\documentclass{article}
\usepackage{acra}
\usepackage{url}
\usepackage{graphicx}
\usepackage{subcaption}
\usepackage{adjustbox}

\title{An Ontology-driven Dynamic Knowledge Base for Uninhabited Ground Vehicles}
\author{
\textbf{Hsan Sandar Win$^{1*}$, Andrew Walters$^{2}$, Cheng-Chew Lim$^{1}$}\\
\textbf{Daniel Webber$^{1}$, Seth Leslie$^{2}$, Tan Doan$^{2}$} \\
$^{1}$The University of Adelaide, SA, Australia \\
$^{2}$Defence Science and Technology Group, Edinburgh, SA, Australia \\
hsansandar.win@adelaide.edu.au$^{1*}$
}

\begin{document}

\maketitle

\begin{abstract}
In this paper, the concept of Dynamic Contextual Mission Data (DCMD) is introduced to develop an ontology-driven dynamic knowledge base for Uninhabited Ground Vehicles (UGVs) at the tactical edge. The dynamic knowledge base with DCMD is added to the UGVs to: support enhanced situation awareness; improve autonomous decision making; and facilitate agility within complex and dynamic environments. As UGVs are heavily reliant on the \textit{a priori} information added pre-mission, unexpected occurrences during a mission can cause identification ambiguities and require increased levels of user input. Updating this \textit{a priori} information with contextual information can help UGVs realise their full potential. To address this, the dynamic knowledge base was designed using an ontology-driven representation, supported by near real-time information acquisition and analysis, to provide in-mission on-platform DCMD updates. This was implemented on a team of four UGVs that executed a laboratory based surveillance mission. The results showed that the ontology-driven dynamic representation of the UGV operational environment was machine actionable, producing contextual information to support a successful and timely mission, and contributed directly to the situation awareness.
\end{abstract}

\section{Introduction}
Uninhabited Ground Vehicles (UGV) are increasingly deployed in complex and dynamic environments to lessen the physical and cognitive burden of human operators \cite{Army} at the tactical edge. In such settings, situation awareness (SA) enables decision-making to support mission success, providing knowledge derived from a complete understanding of a situation \cite{DRI} for UGVs. SA requires timely access to relevant and up-to-date information \cite{10017459,DRI} to facilitate decision making.

Since UGVs rely on static \textit{a priori} information, their ability to adapt to new situations in the environment is often curtailed. This limitation can lead  to uncertainties in SA, decision inefficiencies and continued dependence on human UGV operators \cite{badue2021self}. 
In contrast, the real-time multimodal data from UGVs provides timely observations but lacks the semantic depths required to distinguish the merely observable from the relevant context to the situation. 

In order to provide the ``useful input to a situational picture" \cite{8170353}, the real-time multimodal data, both structured and unstructured, could be leveraged to provide in-mission on-platform updates to the UGV's knowledge base \cite{DRI,sun2019high} by integrating with \textit{a priori} information. As such the concept of Dynamic Contextual Mission Data (DCMD) is reported here. Building on the formal description of contextual and actionable information in \cite{DRI}, this work details the implementation of a DCMD concept to design an ontology-driven dynamic knowledge base for updating \textit{a priori} information from UGV computer vision outputs. 

The ontology-driven knowledge base provides a structured approach to formalising and representing the DCMD updates, enabling the representation of domain knowledge and relationships between objects \cite{SAWA} in UGV systems.
Moreover, the ontology-driven knowledge base supports semantic interoperability \cite{interoperable} among the UGVs to share DCMD updates to achieve a common understanding of a situation and to perform coordinated actions. 

The proposed dynamic knowledge base adopted upper-level ontologies to support interoperability as they provided the ``most general domain-independent categories of reality" including time and space, individuals, objects, events, process and instantiation'' \cite{upperlevel_ontologies}.  Specifically, the Basic Formal Ontology (BFO) was chosen for the upper-level ontology \cite{BFO} and Common Core Ontologies (CCO) was utilised as a mid-level semantic layer that supported consistent representation and integration of mission-relevant concepts \cite{CCO}. Using these ontologies with the DCMD concept, the main contributions reported in this paper are:
\begin{itemize}
	\item An ontology-driven representation to model the DCMD concept within a knowledge graph.
	\item An on-board knowledge base deployed on a multi-agent team, integrating with near real-time information acquisition, analysis and dynamic knowledge graph updates to support situation awareness for UGVs at the tactical edge.
\end{itemize}
This paper describes the DCMD concept and its implementation for the dynamic knowledge base in Section 2. Section 3 presents the experimental setup and Section 4 details experimental results and discussion for DCMD updates within a laboratory-based surveillance mission. Section 5 presents conclusions and future work.

\section{DCMD Concept} \label{DCMD_concept}
\begin{figure}[!htbp]
	\centering
	\includegraphics[width=\columnwidth]{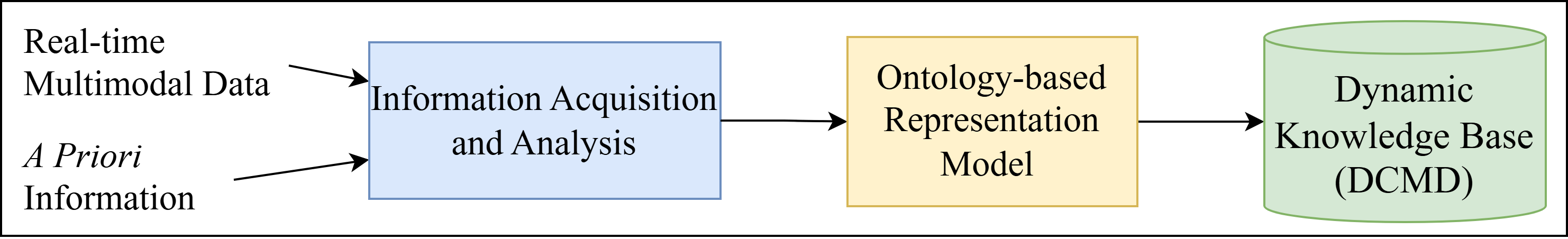}
	\caption{DCMD concept}
	\label{fig:concept}
\end{figure}

DCMD builds on \textit{a priori} information, exploiting real-time multimodal data to generate a dynamic knowledge base. It contains information that is contextual to the mission and immediately actionable towards mission goals. The DCMD concept is presented in Fig.~\ref{fig:concept}. 

Both multimodal data and \textit{a priori} information feed into the information acquisition and analysis process to identify and prepare information for the representation stage. The results give the information context within the current mission; covering meaning, relevance, and usefulness. The resultant dynamic knowledge base contains information that adds a mission-specific spatial-temporal context to the \textit{a priori} information, which is decision-ready in response to unexpected mission events or outcomes.

In order to test the DCMD concept, an implementation was designed for a team of four UGVs conducting a surveillance mission. The scenario details for the mission were as follows: (1) An intelligence report indicated that an airfield faced a potential hazard concealed within a nearby town (2) The objective of the mission was to locate and verify this hazard before it could do any harm on the airfield. To achieve the objective, the agents were divided into two teams:
\begin{enumerate} 

	\item Explorer Team: Tasked with scanning the town, confirming the positions of objects of interest, and identifying hazards. They collected information at predefined waypoints (WPs) in an operational area and generated DCMD updates, which were shared among the UGVs according to their roles.
	\item Verifier Team: Activated after hazard detection by the Explorer team. They navigated to target locations and verified the detected hazards.

\end{enumerate}

\begin{figure*}[htbp]
	\centering
	\includegraphics[width=0.9\textwidth]{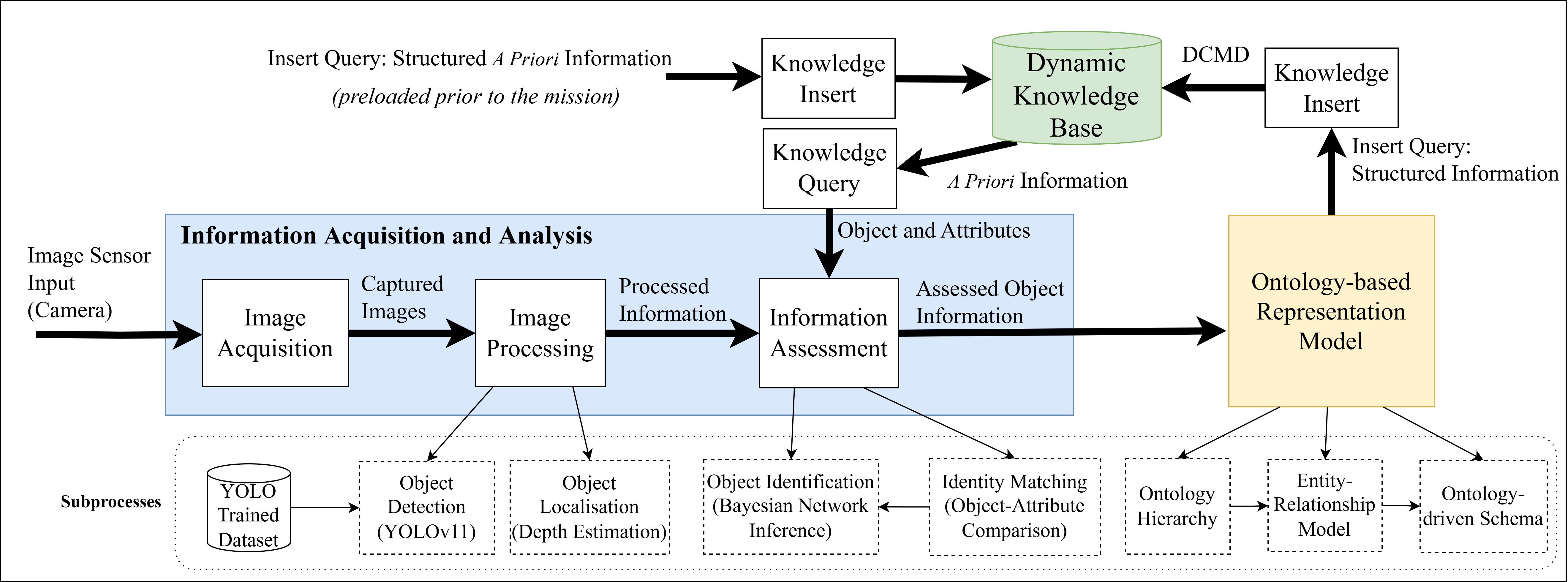}
	\caption{On-platform DCMD concept implementation for UGVs mission}
	\label{fig:model}
\end{figure*}

Fig.~\ref{fig:model} presents the detailed structure of the DCMD implementation. The information required for the mission contained objects of interest, their associated attributes, and environmental attributes. Based on this information, the implementation used object-centric processes
in which images serve as the primary input within a real-time multimodal data stream.   

\subsection{Information Acquisition and Analysis}
Information acquisition and analysis shown in Fig.~\ref{fig:model} involved the following main processes:
\subsubsection{Image Aquisition} Sensed and captured visual data for subsequent processing. 
\subsubsection{Image Processing} Performed object detection using the YOLOv11 for an object's class and its attributes; and object localisation using depth estimation for object location. 
\subsubsection{Information Assessment} Determined an identity of the processed object using (1) an identity matching algorithm and (2) Bayesian Network (BN) inference to provide its probability of being known object (non-hazard) or unknown object (hazard). The identity matching algorithm queried known objects from the knowledge base and compared them with the processed object. When a match occurred, the corresponding attributes were compared. The resultant object and attribute match statuses along with their confidence levels were compiled into an evidence set and passed into the BN inference model. 

The BN inference model determined whether the processed object was already known in the \textit{a priori} information or if it was new, and provided the posterior probability output. If the object was identified as new, the BN inference evaluated its hazard level using additional associated objects (weapon and person) extracted from the object detection stage. The result of the inference was passed into the representation model along with its attributes. The evidence block contained the nodes representing the evidence sets from Identity matching, and the hypothesis node represents the query variable of object identity, which was an outcome of the BN inference. 

\begin{figure}[!htbp]
	\centering
	\includegraphics[width=0.9\columnwidth]{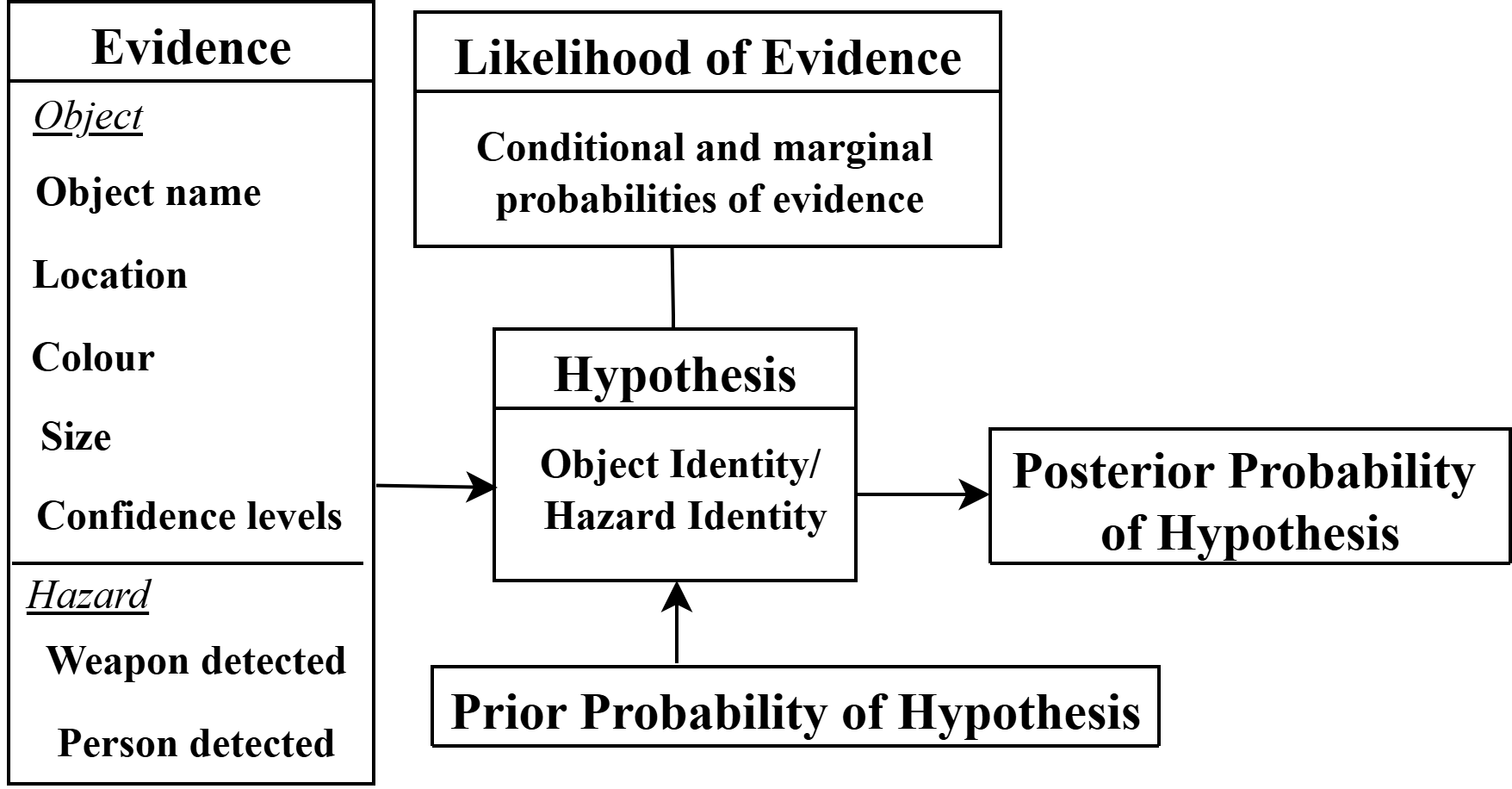}
	\caption{Bayesian Network inference model for object and hazard identification.}
	\label{fig:BN}
\end{figure}

The implementation of BN for object and hazard identification can be seen in Fig.~\ref{fig:BN}. This BN used discrete-valued state for the evidence set. Using discrete variables simplifies the structure and computations involved in probabilistic inference. The \textit{a priori} marginal probabilities and conditional probabilities were populated for both the evidence set and the hypothesis. 

The BN inference model used Variable Elimination (VE) for exact inferencing. It eliminated those variables irrelevant to the query one by one \cite{rollon2015variable}. VE reduced the computational burden by focusing on relevant variables and efficiently handling interdependencies in a BN \cite{rollon2015variable}. By enabling real-time inference under limited computational resources, VE supported the implementation of a dynamic knowledge base that met the requirements of operating on-platform and in-mission.
 
\subsection{Ontology-based Representation Model}
The representation model combined ontology hierarchy and an entity-relationship (ER) model to develop an ontology-driven schema for the knowledge base. 
\subsubsection{Ontology Hierarchy}
The ontology hierarchy adopted BFO as an upper-level ontology for its realist approach to represent the world objectively, rather than depending on natural language and human common sense \cite{upperlevel_ontologies}. It provided a universal, high-level semantic framework under which all domain concepts were unified \cite{BFO} unlike other upper-level ontologies which are too descriptive and less realistic, and also not sufficiently universal \cite{upperlevel_ontologies}. 

As the mid-level layer of the ontology hierarchy, the CCO represented specific entities and relationships relevant to mission-centric domains. CCO provided a mid-level semantic layer that supported consistent representation and integration of mission-relevant concepts \cite{CCO}. CCO enforces semantic categories throughout the hierarchy of entities, and reduces the need for a customised ontology at the domain level \cite{CCO} to foster knowledge sharing and updating across the UGV team. To provide semantic definitions of relationships among the entities defined by BFO and CCO, the Relation Ontology Core (RO Core) was also integrated into the ontology hierarchy, providing domain-neutral relations \cite{ro-core}.

While BFO, CCO and RO Core provided a reusable semantic backbone, the domain-level ontology was designed to provide structured and semantically rich representations for dynamic operational contexts specific to the UGVs surveillance mission. It considered objects of interest, agents, and operational forces, and their associated operational areas. The ontology hierarchy served as the conceptual layer of the representation model while the ER model served as the logical layer to map the entities and relationships into the schema. 
\begin{figure*}[!htbp]
	\centering
	\includegraphics[width=0.79\textwidth]{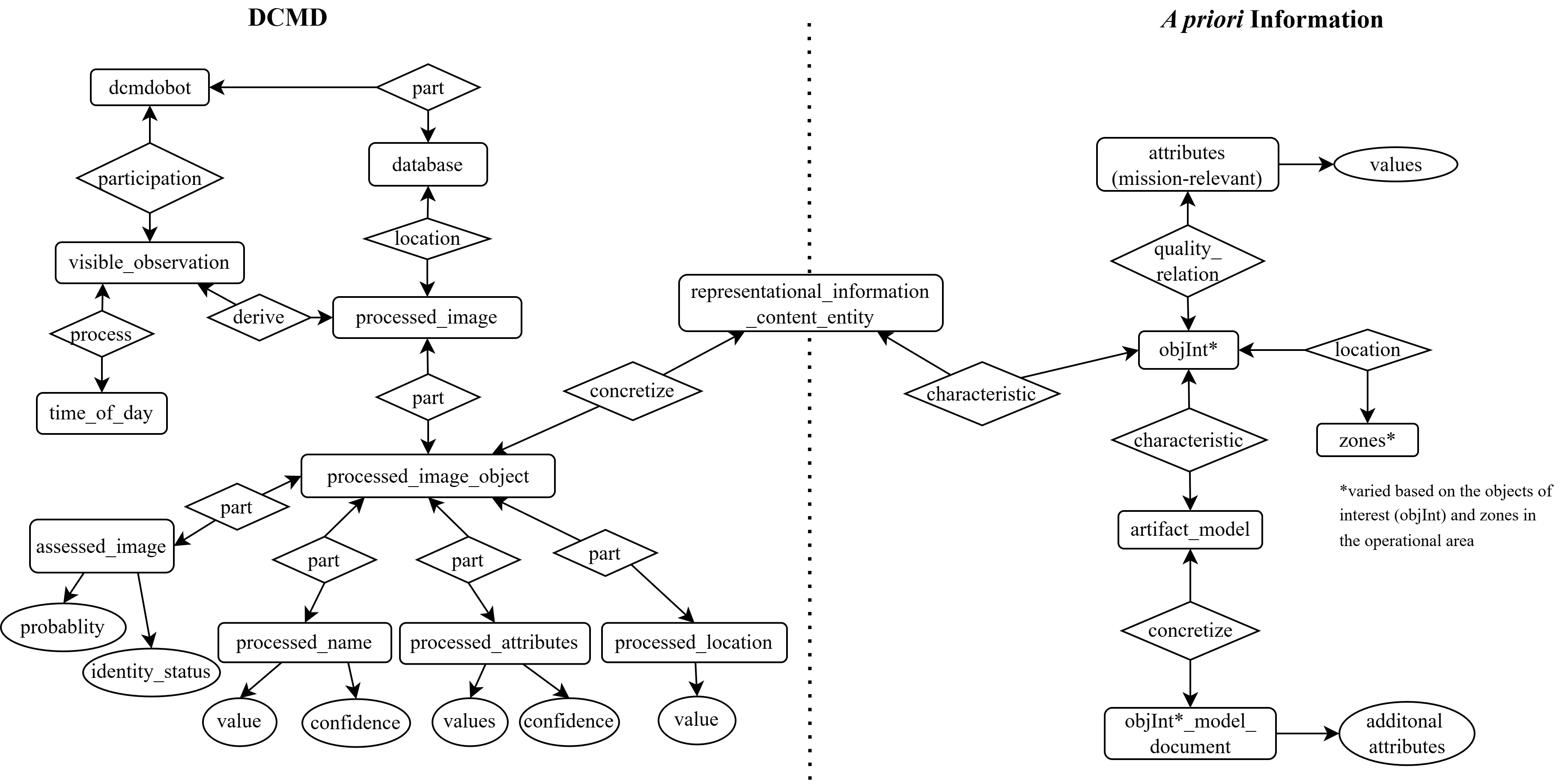}
	\caption{Subset of the overall schema for objects of interest identified (rectangles for entities, ellipses for attributes, and diamonds for relations).}
	\label{fig:object_schema}
\end{figure*}

\subsubsection{Entity-Relationship Model}
The ER model is a semantic data modeling technique, used to define the meaning of data and represent the real world symbolically within physical data stores \cite{Li2009}. It visually represents entities, their attributes and relationships, serving as the basic structure of the schema. At the logical and physical level, the ontology concepts were mapped into the ER model to form the logical schema for the implementation of the knowledge base in a database system. In doing so, the schema reflected both the semantic richness of the ontologies and the structural coherence of the ER model. Fig.~\ref{fig:object_schema} presents a subset of the overall schema, highlighting how DCMD updates of objects of interest were represented in the knowledge base.

The schema structure shown in Fig.~\ref{fig:object_schema} is divided into two parts, representing \textit{a priori} information and DCMD. In the \textit{a priori} information part, objects were characterised by the {\small\texttt{artifact\_model}} entity with an object-related document and linked via {\small\texttt{quality\allowbreak\_relation}} to mission-relevant attributes. If an object was not listed as part of the known objects, it was represented only by the general class described by the {\small\texttt{artifact\_model}} entity. In the DCMD part of Fig.~\ref{fig:object_schema}, the representation varied depending on the identity of the detected object. It included updates of the object's spatial and temporal information, source of information, and its detected attributes. The connection between DCMD updates and the \textit{a priori} information was established via the {\small\texttt{representational\allowbreak\_information\allowbreak\_content}} entity.

\subsection{Dynamic Knowledge Base}
The knowledge base required a database that could naturally express ontological structures and support reasoning. TypeDB, a polymorphic database with a conceptual data model \cite{typedb}, was chosen for the on-board knowledge base, as it provided the capabilities of a knowledge graph together with the formality of an ontology. TypeDB defines entity types, relation types, and attributes that align with the structure of the ER model, along with type hierarchies and role constraints. Moreover, BFO and CCO entities can be instantiated as TypeDB entity types, and RO Core relations can be modeled as TypeDB relations with roles. 

The Knowledge Query and Insert interfaces shown in Fig.~\ref{fig:model} were implemented using TypeDB Query Language (TypeQL). There are two categories for queries in a TypeDB database: 1) schema queries; and 2) data pipelines \cite{typedb}. This implementation used data pipeline queries to read and write DCMD updates to the knowledge base during the mission. Prior to the mission, the \textit{a priori} information including mission contexts and known objects of interest (non-hazard) was loaded to each agent's knowledge base using the Knowledge Insert interface. The DCMD updates were shared among the team based on the type of information collected and the roles of the agents. 

\begin{figure}[!htbp]
	\centering
	\includegraphics[width=0.8\columnwidth]{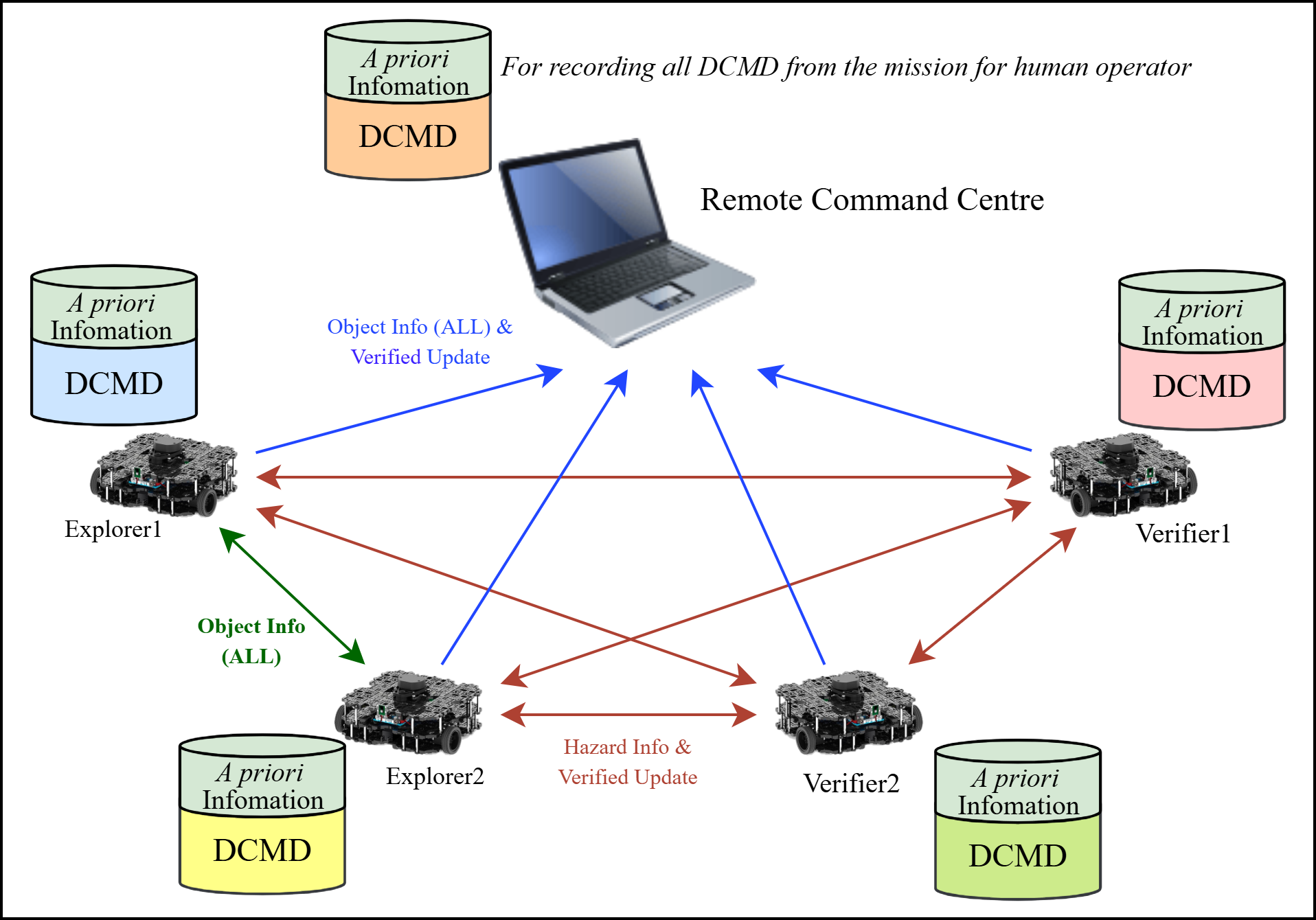}
	\caption{On-board dynamic knowledge bases and their information flows among a team of four UGVs and RCC. The communication among the agents used a Wireless Local Area Network (WLAN) architecture, in which each agent and the RCC wirelessly connect to a central Wireless Access Point (WAP). Color coding of the DCMD in each knowledge base reflects the role-specific contextual information associated with each agent.}
	\label{fig:info_flow}
\end{figure}

The setup of the dynamic knowledge bases and their information flows among teams of UGVs are illustrated in Fig.~\ref{fig:info_flow}. The Remote Command Centre (RCC) was considered as a human operator who only monitored the mission progress through information shared from the UGVs, therefore it participated in the mission as a passive observer.

\section{Experimental Setup}
\begin{figure}[!htbp]
	\centering
	\begin{subfigure}[b]{0.9\columnwidth}
		\centering
		\includegraphics[width=\textwidth]{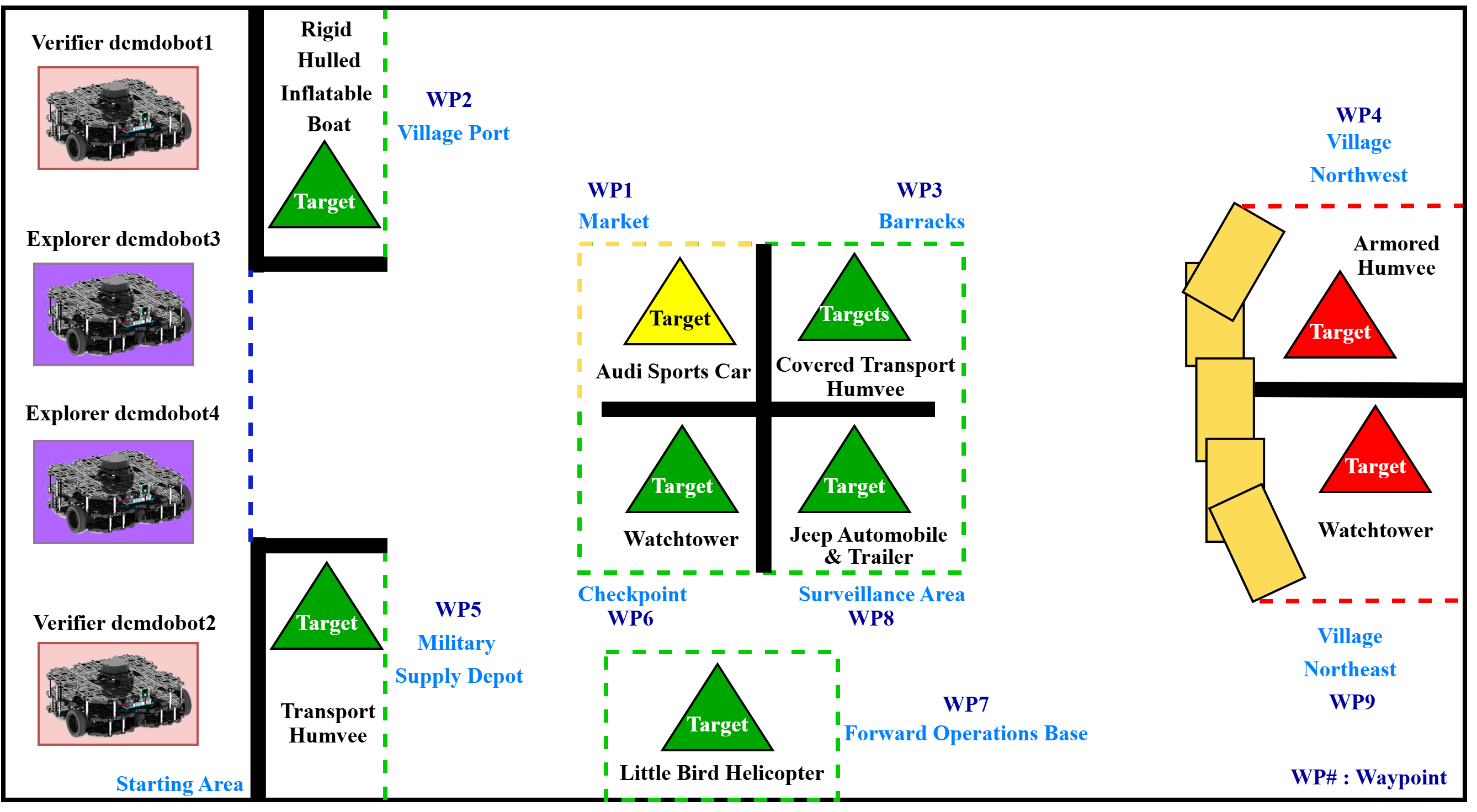}
		\caption{Schematic layout}
		\label{fig:fig1}
	\end{subfigure}
	\begin{subfigure}[b]{0.9\columnwidth}
		\centering
		\includegraphics[width=\textwidth]{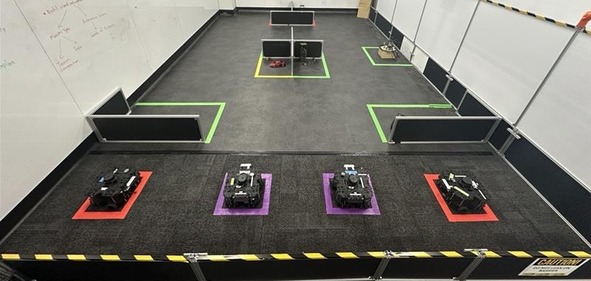}
		\caption{Physical testbed}
		\label{fig:fig2}
	\end{subfigure}
	\caption{Operational area configured for the scenario. Green targets are known objects, red targets are hazards, and the yellow targets are civilian.}
	\label{fig:testbed}
\end{figure}

A physical testbed was constructed to facilitate experimentation representing the town near the airfield mentioned in the scenario in Section~\ref{DCMD_concept}. A layout of the operational area is shown in Fig.~\ref{fig:testbed}. It includes a schematic layout and an overhead photograph of the physical testbed. The testbed had dimensions of \ensuremath{6 \times 2\, \mathrm{m}} and contained $1\!:\!18$ scale plastic target models, providing a representative environment for UGV operations. 

Four UGV agents were assigned as follows: \textit{dcmdobot1} and \textit{dcmdobot2} in Verifier team; and \textit{dcmdobot3} and \textit{dcmdobot4} in Explorer team. The hardware platform used for the agents was the TurtleBot3 Waffle-Pi from Robotis \cite{robotis_website}, equipped with a Nvidia Jetson ORIN AGX \cite{nvidia2022jetson}, which included a Graphics Processing Unit (GPU) for enhanced on-board computer vision processing performance. Their primary sensor for the experiments was the RealSense D435i camera for object detection and the LiDAR and IMU sensors for agent localisation, navigation and obstacle avoidance. The overall operation architecture was the Robot Operating System 2 (ROS 2) \cite{ros2humble2022} for managing robotic operations (navigation and path planning) and the DCMD concept implementation for populating the knowledge base.

\section{Results and Discussion}
\begin{figure*}[!ht]
  \centering

  \begin{subfigure}[b]{0.3\textwidth}
    \includegraphics[height=4.5cm, width=\textwidth]{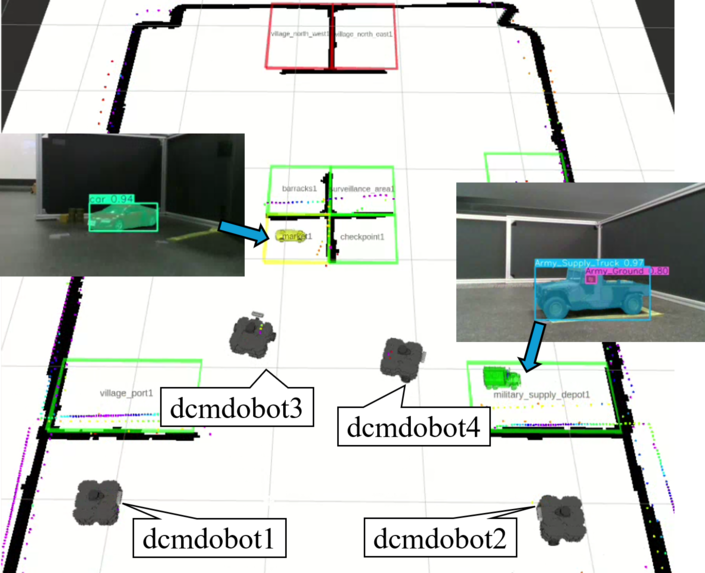}
    \caption{Explorer team identified known objects of interest}
    \label{fig:fig1_op}
  \end{subfigure}
  \hfill
  \begin{subfigure}[b]{0.3\textwidth}
    \includegraphics[height=4.8cm, width=\textwidth]{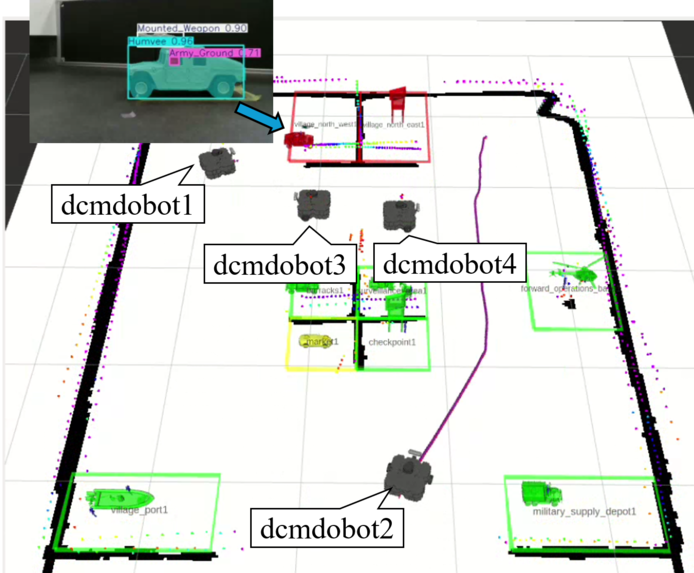}
    \caption{\textit{dcmdobot1} verified hazard}
    \label{fig:fig2_op}
  \end{subfigure}
  \hfill
  \begin{subfigure}[b]{0.3\textwidth}
    \includegraphics[height=4.72cm, width=\textwidth]{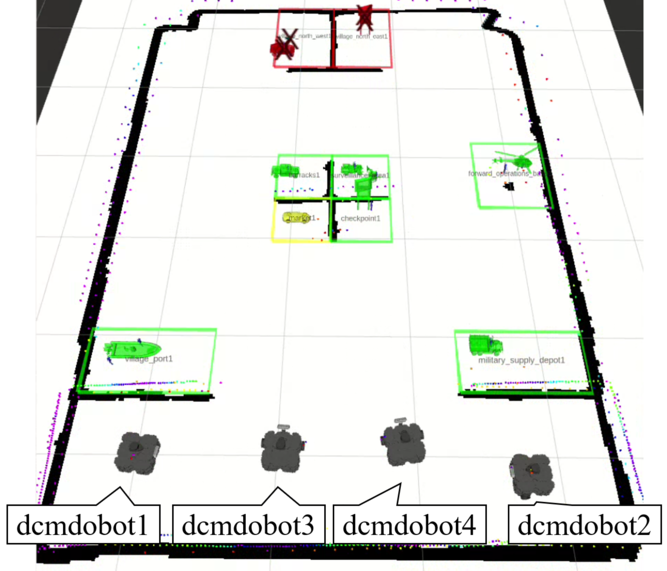}
    \caption{Mission complete}
    \label{fig:fig3_op}
  \end{subfigure}

  \vskip1em   
  \begin{subfigure}[b]{0.45\textwidth}
    \includegraphics[height=4.5cm,width=\textwidth]{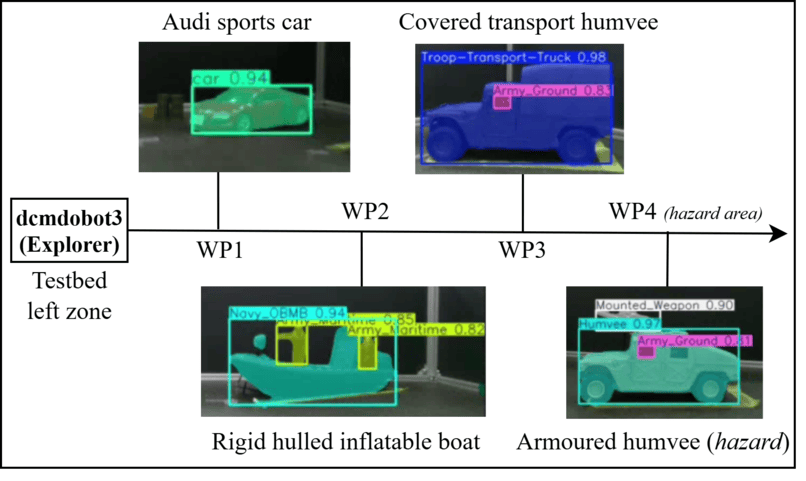}
    \caption{\textit{dcmdobot3}'s detections}
     \label{fig:fig1_d}
  \end{subfigure}
  \hfill
  \begin{subfigure}[b]{0.45\textwidth}
    \includegraphics[height=4.55cm,width=\textwidth]{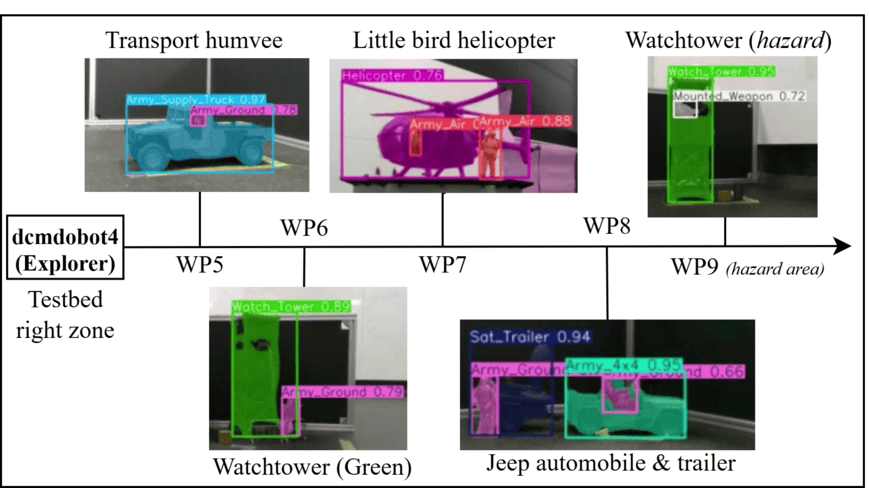}
    \caption{\textit{dcmdobot4}'s detections}
     \label{fig:fig1_e}
  \end{subfigure}

  \caption{Example operational stages and detection results during the mission}
  \label{fig:mission}
\end{figure*}

Experimental runs were conducted in which the UGV agents executed their mission. Multiple runs were conducted to assess the execution repeatability. A typical experimental run was selected to analyse and demonstrate how DCMD updates were generated and represented in the knowledge base during the mission. The generated DCMD updates triggered a sequence of events and actions, realised through the cooperative coordination of the Explorer and Verifier. Fig.~\ref{fig:mission} shows the operational stages and detection results of a successful surveillance mission, which provided the common operating picture.

Fig.~\ref{fig:fig1_op} illustrates the events executed by \textit{dcmdobot3} and \textit{dcmdobot4}, in which they started scanning the area and confirming the location of known objects of interest. Fig.~\ref{fig:fig2_op} shows the event triggered when \textit{dcmdobot1} scanned the \texttt{village\allowbreak\_northwest} area to verify the hazard, following the detection of an armed \texttt{armored\allowbreak\_humvee} by \textit{dcmdobot3}. Moreover, it shows that \textit{dcmdobot2} was also on the way to verify \texttt{watchtower}, which was detected by \textit{dcmdobot4} in the \texttt{village\allowbreak\_northeast} area.

The successful completion of the mission is shown in Fig.~\ref{fig:fig3_op}, in which all hazards were identified by the Explorer team and verified by the Verifier team. In additon, all known objects of interest were successfully identified at their designated locations by the Explorer team. Fig.~\ref{fig:fig1_d} and Fig.~\ref{fig:fig1_e} show the events of the Explorer team, including detection of objects at each waypoint, and identification of the hazards that triggered the deployment of the Verifier team. 

The following analysis of the experimental results examines whether the DCMD implementation successfully achieved a dynamic knowledge base for UGV operations.

\subsubsection{\textbf{DCMD Updates of Known Object of Interest}}
The detection of a \textit{boat} and its assigned people by \textit{dcmdobot3} occurred at the {\small\texttt{village\allowbreak\_port}} (WP2), as illustrated in Fig.~\ref{fig:fig1_d}. The information extracted from the detection through the information acquisition and analysis process during the experimental run is listed in Table~\ref{tab:processed} and Table~\ref{tab:assessed}.

\begin{figure*}[!ht] 
	\centering
	\includegraphics[width=\textwidth]{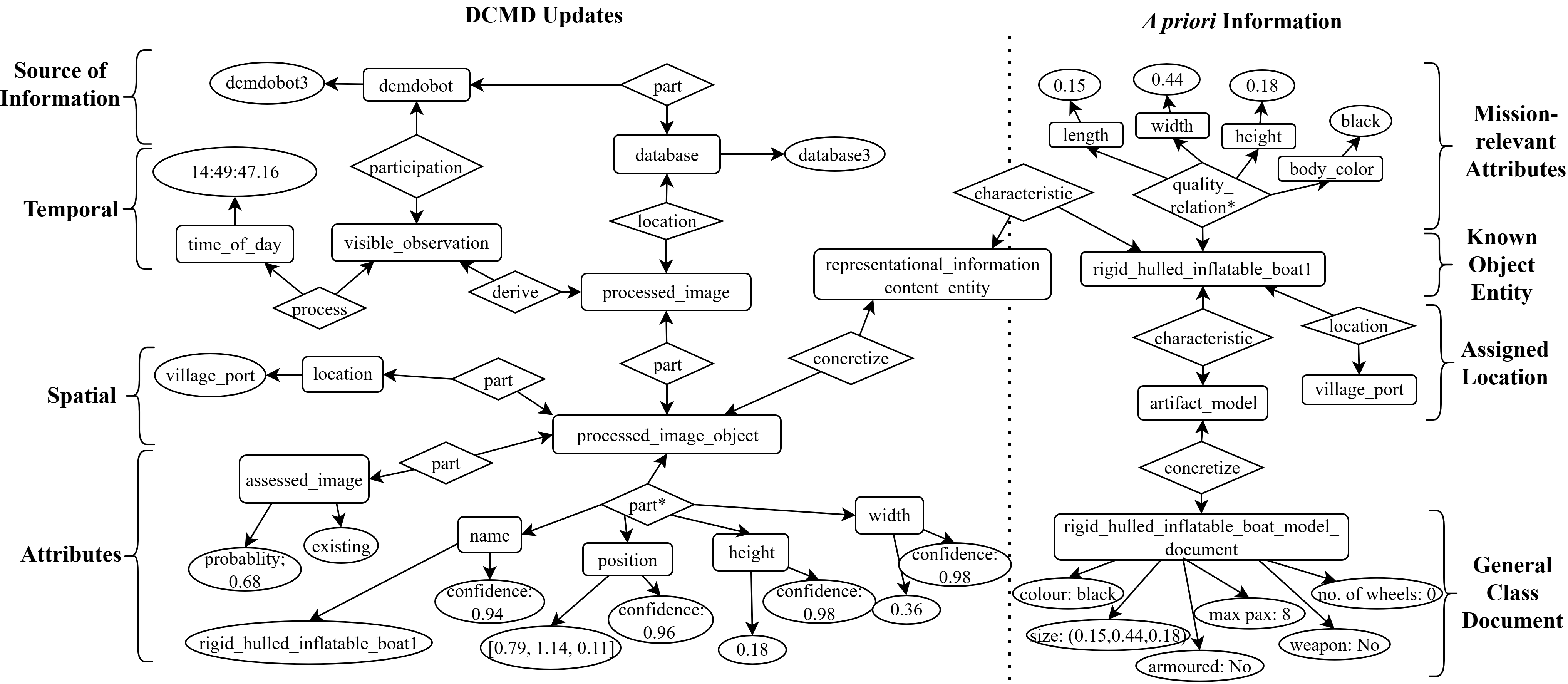}
	\caption{DCMD updates of the {\small\texttt{rigid\allowbreak\_hulled\allowbreak\_inflatable\allowbreak\_boat}} represented using the ontology-driven schema in the dynamic knowledge base (TypeDB)}
	\label{fig:boat_rep}
\end{figure*}

Table~\ref{tab:processed} contains three detected objects including the \textit{boat} and two \textit{army\allowbreak\_maritime} personnel, located in the {\small\texttt{village\allowbreak\_port}}. For each object, the processed information contained its attributes of object name, position, size (height and width), and their corresponding confidence levels (CL). Table~\ref{tab:assessed} shows that the \textit{boat} was identified as a {\small\texttt{rigid\allowbreak\_hulled\allowbreak\_inflatable\allowbreak\_boat}} with a probability of 0.68 and two \textit{army\allowbreak\_maritime} were as {\small\texttt{known\allowbreak\_person}} with probabilities of 0.73 and 0.74, respectively. Therefore, these objects were inferred as known objects and were not identified as hazards. This information was subsequently represented in the knowledge base, where the objects were updated with their location, detection time, and identity attributes; connecting with the \textit{a priori} records of these objects. The representation of those updates in the knowledge base is shown in Fig.~\ref{fig:boat_rep}.

\begin{table}[htbp]
  \centering
  \caption{Processed information extracted from \textit{boat} at 14:49:47.16 in the {\small\texttt{village\_port}}}
  \label{tab:processed}
  \begin{adjustbox}{width=\columnwidth}
    \begin{tabular}{|c|c|c|c|c|c|c|}
      \hline
      \textbf{obj\_name} & \textbf{position} & \textbf{height} & \textbf{width} & 
      \textbf{obj\_CL} & \textbf{position\_CL} & \textbf{size\_CL} \\
      \hline
      boat & 0.79, 1.14, 0.11 & 0.18 & 0.36 & 0.94 & 0.96 & 0.98 \\
      \hline
      army\_maritime & 0.85, 1.09, 0.15 & 0.08 & 0.05 & 0.79 & 0.87 & 0.95 \\
      \hline
      army\_maritime & 0.69, 1.29, 0.13 & 0.08 & 0.04 & 0.82 & 0.96 & 0.98 \\
      \hline
    \end{tabular}
  \end{adjustbox}
\end{table}
\begin{table}[htbp]
  \centering
  \caption{Assessed information from \textit{boat} at 14:49:47.16 in the {\small\texttt{village\_port}}}
  \label{tab:assessed}
  \begin{adjustbox}{width=\columnwidth}
    \begin{tabular}{|c|c|c|}
      \hline
      \textbf{obj\_name} & \textbf{identity} & \textbf{probability} \\
      \hline
      boat & rigid\_hulled\_inflatable\_boat1 & 0.68 \\
      \hline
      army\_maritime & known\_person10 & 0.73 \\
      \hline
      army\_maritime & known\_person11 & 0.74 \\
      \hline
    \end{tabular}
  \end{adjustbox}
\end{table}

\begin{figure*}[!t] 
	\centering
	\includegraphics[width=0.9\textwidth]{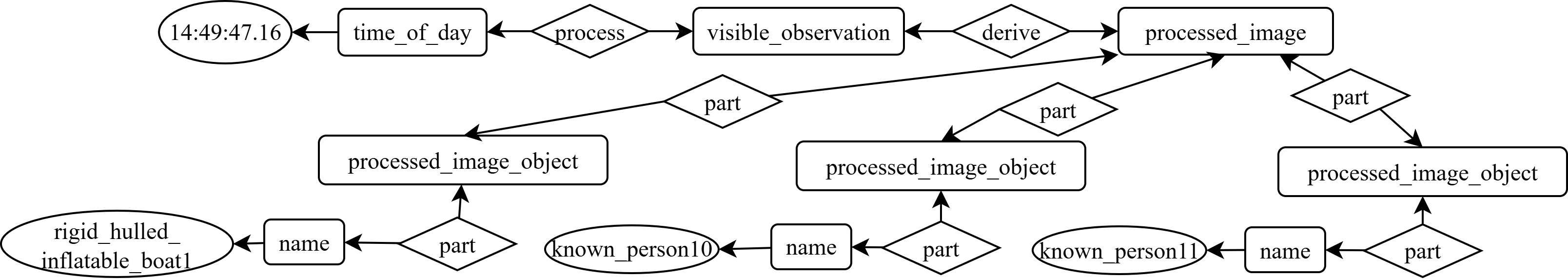}
	\caption{Representation of the three detected objects: {\small\texttt{rigid\allowbreak\_hulled\allowbreak\_inflatable\allowbreak\_boat}}, {\small\texttt{known\allowbreak\_person10}}, {\small\texttt{known\allowbreak\_person11}} at 14:49:47.16, showing the relationships formed between them in the dynamic knowledge base.} 
	\label{fig:objects_rep}
\end{figure*}

The {\small\texttt{part*}} and {\small\texttt{quality\allowbreak\_relation*}} shown in Fig.~\ref{fig:boat_rep} for attributes are visually grouped for display purposes. In the underlying representation, each attribute maintains its own relation. The representation  shows how the DCMD updates of the detected \textit{boat} were semantically represented using the ontology-driven schema, linking to the \textit{a priori} information record of the {\small\texttt{rigid\allowbreak\_hulled\allowbreak\_inflatable\allowbreak\_boat}} in the knowledge base. The DCMD representation reflected the real-time updates of the {\small\texttt{rigid\allowbreak\_hulled\allowbreak\_inflatable\allowbreak\_boat}} during the mission. The DCMD updates were then linked to the \textit{a priori} record which contained the mission-relevant attributes, the assigned location and the general class document detailing relevant characteristics of the {\small\texttt{rigid\allowbreak\_hulled\allowbreak\_inflatable\allowbreak\_boat}}. The resultant representation demonstrates how the real-time information was semantically integrated with the \textit{a priori} information, and updated within the dynamic knowledge base during the mission.

Following the representation of the {\small\texttt{rigid\allowbreak\_hulled\allowbreak\_inflatable\allowbreak\_boat}}, Fig.~\ref{fig:objects_rep} shows how it formed relationships with other objects ({\small\texttt{known\allowbreak\_person10}} and {\small\texttt{known\allowbreak\_person11}}), detected at 14:49:47.16 in the {\small\texttt{village\allowbreak\_port}}. The representations of the known personnel followed the same structure as that of the {\small\texttt{rigid\allowbreak\_hulled\allowbreak\_inflatable\allowbreak\_boat}} shown in Fig.~\ref{fig:boat_rep}. Each {\small\texttt{processed\allowbreak\_image\allowbreak\_object}} entity for each object was linked directly to the {\small\texttt{processed\allowbreak\_image}} entity, capturing the contextual description of the detection event. This result shows that the relationship representation of the detected objects was successfully integrated into the knowledge base, enabling the agents to infer contextual meaning about the detection event as a whole, beyond isolated objects.

\subsubsection{\textbf{DCMD Updates of the Hazard}}
As the mission progressed, the Explorer agents explored and updated their knowledge bases with DCMD updates of the known objects of interest, sharing them among themselves. Hazards were detected as they advanced toward the north area of the village. At 14:50:44.67, \textit{dcmdobot3} identified an unknown \textit{humvee} in {\small\texttt{village\allowbreak\_northwest}} (WP4), and at 14:51:13.27, \textit{dcmdobot4} detected an unknown \textit{watchtower} in {\small\texttt{village\allowbreak\_northeast}} (WP9) (see Fig.~\ref{fig:fig1_d}). 

The \textit{humvee} was detected with unknown \textit{army\allowbreak\_ground} person and a \textit{mounted\allowbreak\_weapon} in {\small\texttt{village\allowbreak\_northwest}}. For each object, the processed information contained its attributes and their corresponding confidence levels. The assessed information showed the identity of the unknown object linking with the general class documents of {\small\texttt{armoured\allowbreak\_humvee}} for the \textit{humvee} and {\small\texttt{mk19\allowbreak\_grenade\allowbreak\_launcher}} for the \textit{mounted\_weapon}. The unknown \textit{army\allowbreak\_ground} was identified as {\small\texttt{hazard\allowbreak\_related\allowbreak\_person1}} and linked to the general class of  {\small\texttt{army\allowbreak\_ground}} stored in the \textit{a priori} information. 
\begin{figure*}[!ht] 
	\centering
	\includegraphics[width=\textwidth]{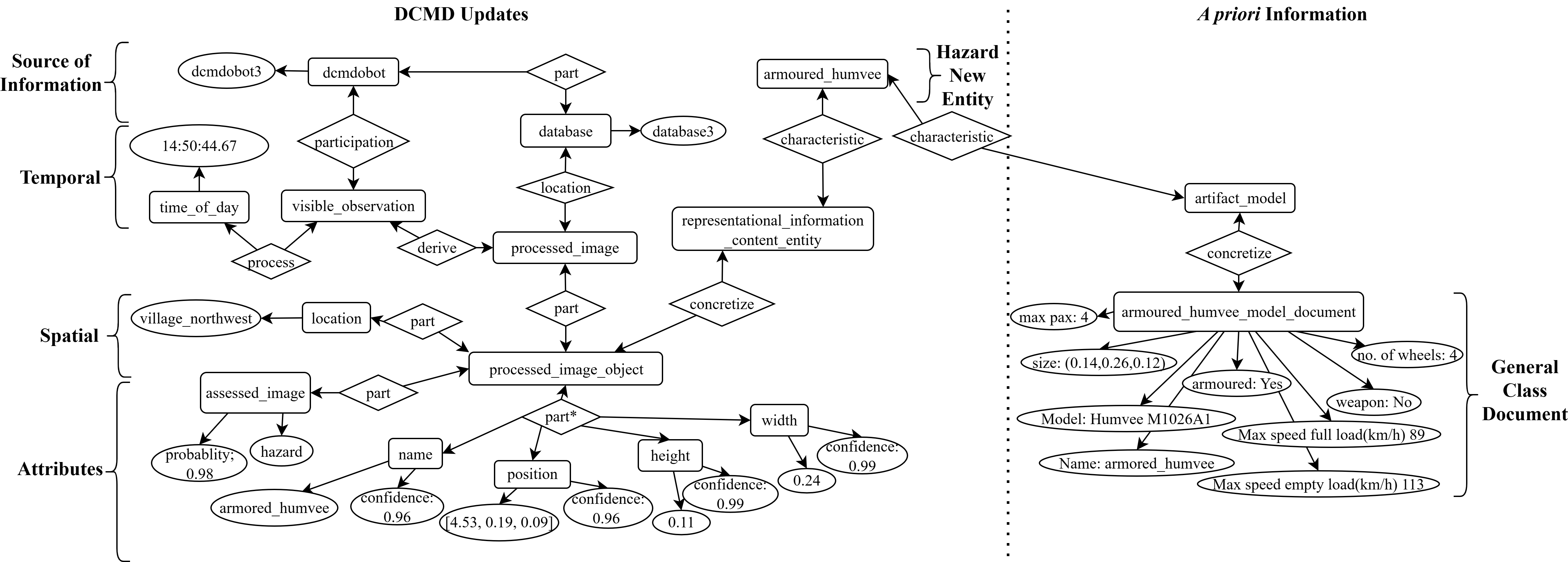}
	\caption{DCMD updates of the {\small\texttt{armoured\allowbreak\_humvee}} ontology-based representation in the dynamic knowledge base.}
	\label{fig:humvee_rep}
\end{figure*}

\begin{figure*}[!ht] 
	\centering
	\includegraphics[width=\textwidth]{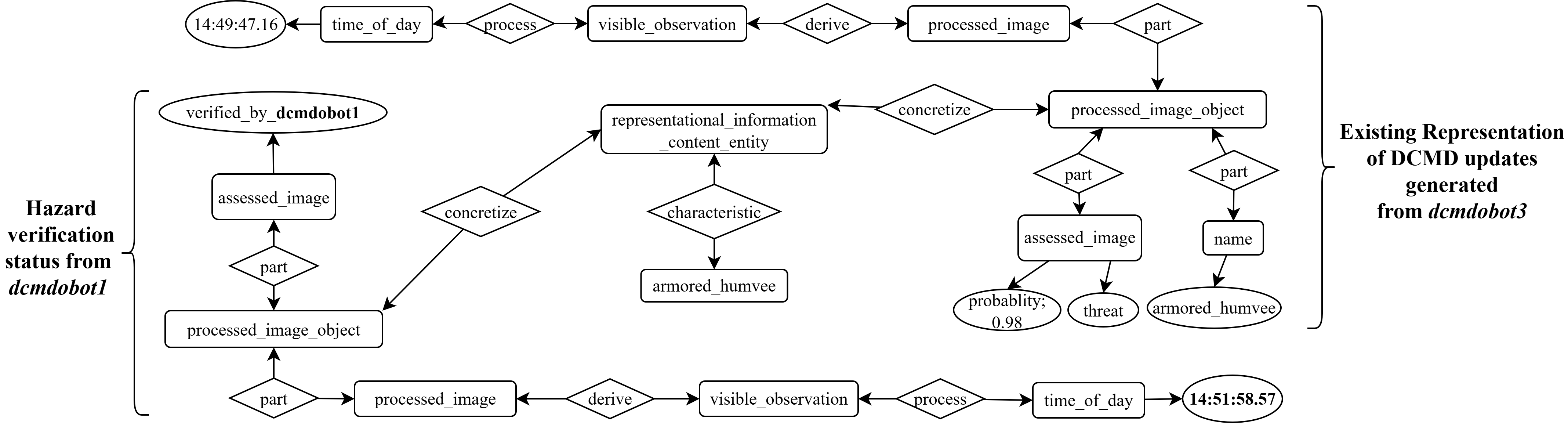}
	\caption{Integration of the updated hazard status for the {\small\texttt{armoured\allowbreak\_humvee}} into the \textit{dcmdobot3}'s knowledge base from the \textit{dcmdobot1}}
	\label{fig:threat_status_humvee}
\end{figure*}

The implementation of assessing what constituted a hazard considered the presence of a weapon and a person to operate it. As a result, the detected \textit{humvee} was identified as a hazard with a probability of 0.98. Its representation updated in the knowledge base as a hazard and can be seen in Fig.~\ref{fig:humvee_rep}. Similar to the representation of the known object, DCMD updates of the hazard {\small\texttt{armoured\allowbreak\_humvee}} contained spatial and temporal information, source of information and its detected attributes, as shown on the left side of Fig.~\ref{fig:humvee_rep}. In addition, the hazard identity and its probability were also recorded in the representation. Since the temporal spatial version of this hazard was not predefined in the \textit{a priori} information, the new entity called {\small\texttt{armoured\allowbreak\_humvee}} was created in the update to provide the connection to its general class document stored in the \textit{a priori} information. Unlike known objects, {\small\texttt{armoured\allowbreak\_humvee}} was only described by its general class in the \textit{a priori} information. 

The relationship representation of the three detected objects in {\small\texttt{village\allowbreak\_northwest}} was subsequently integrated into the knowledge base, allowing agents to infer contextual meaning about the hazard detection event and take action(s). Their relationship representation followed the same structure as that of the detection event shown in Fig.~\ref{fig:objects_rep}, linking each {\small\texttt{processed\_image\_object}} entity to the {\small\texttt{processed\_image}} entity, capturing the contextual description of the hazard detection event that occurred at 14:50:44.67 by \textit{dcmdobot3}. 

The DCMD updates of this hazard detection event in {\small\texttt{village\allowbreak\_northwest}} were shared across the whole team, triggering the Verifier agent \textit{dcmdobot1} to engage and verify all identified hazards. Referring back to the mission overview in Fig.~\ref{fig:mission}, \textit{dcmdobot1}'s action corresponds to the transition into the Verifier operation. At 14:51:58.57, \textit{dcmdobot1} verified {\small\texttt{armoured\allowbreak\_humvee}} in {\small\texttt{village\allowbreak\_northwest}}, followed by the verification of {\small\texttt{army\allowbreak\_ground}} and {\small\texttt{mk19\allowbreak\_grenade\allowbreak\_launcher}}. The \textit{dcmdobot1} agent shared the updates of the verified hazards across the team and each agent updated their knowledge base accordingly. 

The new hazard status of {\small\texttt{armoured\allowbreak\_humvee}} updated in the \textit{dcmdobot3} was selected to display how the new update shared from the \textit{dcmdobot1} was integrated into its existing representation; see Fig.~\ref{fig:threat_status_humvee}. It displays two sets of representations recorded; at 14:50:44.67 and 14:51:58.57 for {\small\texttt{armoured\allowbreak\_humvee}}. The representation recorded at 14:51:58.57 was for the updated hazard status from \textit{dcmdobot1}, displaying the status {\small\texttt{verified\allowbreak\_by\allowbreak\_dcmdobot1}}. This result showed that the DCMD updates from the mission events across the team were successfully captured and integrated into their knowledge bases. 

\section{Conclusion and Future Work}
This paper provided a detailed, scenario-based investigation into the DCMD concept, a dynamic approach to updating a knowledge base in-mission with actionable, contextual information at the tactical edge. The ontology-driven representation model, based on BFO and CCO, was developed to represent objects of interest, their attributes, and the relationship between them and the environment. This ontology-based approach provided a design framework upon which to develop a data structure required to dynamically connect real-time information with \textit{a priori} information; representing the DCMD updates. 

The DCMD implementation was tested and evaluated using a team of four ROS 2-based UGV agents engaged in a surveillance and strike mission. Each UGV was loaded with the \textit{a priori} information and given the task of finding a hazard within the operational area. During the mission, the UGVs identified objects using computer vision and comprehended these objects through real-time analysis. Using information analysis process, all known objects of interest were successfully identified at their assigned locations, whereas hazards were detected in unregistered locations based on the co-occurrence of a weapon and a person capable of operating it.

Identified objects were subsequently updated in the knowledge base using the ontology-driven schema, capturing their inferred identity, locations, attributes, and detection time. These DCMD updates were then linked to the \textit{\textit{a priori}} record which contained the mission-relevant attributes, the assigned location, and general class document detailing relevant characteristics of the identified objects. Objects in the same location and event were semantically connected in the knowledge base, supporting building context between objects and the environment. The DCMD updates were then synchronised across all agents, enabling coordinated actions, such as hazard verification, based on shared updates. The dynamic and context-aware nature of the knowledge base with DCMD contributed directly to the SA, demonstrated through successful completion of the two-phase mission. 

Future research will focus on improving real-time information analysis for assessing the quality of information and addressing uncertainties in the detections in complex situations. Further research is required to investigate the adaptability of the ontology-driven schema and representation of DCMD updates, enabling AI-based adaptive reasoning within the knowledge base to improve SA at the tactical edge.

\bibliographystyle{named}
\bibliography{acra} 




\end{document}